\documentclass[]{tPHM2e}

\begin{document}
\doi{10.1080/14786435.2014.887861}
\issn{1478-6443}
\issnp{1478-6435}
\jvol{00} \jnum{00} \jyear{2014} 

\markboth{Taylor \& Francis and I.T. Consultant}{Philosophical Magazine}

\articletype{Article}

\title{Broken symmetries in URu$_2$Si$_2$}

\author{Takasada Shibauchi$^{\rm a \ast}$\thanks{$^\ast$Corresponding author. Email: shibauchi@k.u-tokyo.ac.jp
\vspace{6pt}}, Hiroaki Ikeda$^{\rm a}$, and Yuji Matsuda$^{\rm a}$\\\vspace{6pt}  $^{\rm a}${\em{Department of Physics, Kyoto University, Sakyo-ku, Kyoto 606-8502, Japan}}\\\vspace{6pt}\received{\today} }

\maketitle

\begin{abstract}
To resolve the nature of the hidden order below 17.5\,K in the heavy fermion compound URu$_2$Si$_2$, identifying which symmetries are broken below the hidden order transition is one of the most important steps. Several recent experiments on the electronic structure have shown that the Fermi surface in the hidden order phase is quite close to the result of band-structure calculations within the framework of itinerant electron picture assuming the antiferromagnetism. This provides strong evidence for the band folding along the c-axis with the ordering vector of $(0\,0\,1)$, corresponding to broken translational symmetry. In addition to this, there is growing evidence for fourfold rotational symmetry breaking in the hidden-order phase from measurements of the in-plane magnetic anisotropy and the effective mass anisotropy in the electronic structure, as well as the orthorhombic lattice distortion. This broken fourfold symmetry gives a stringent constraint that the symmetry of the hidden order parameter should belong to the degenerate $E$-type irreducible representation. We also discuss a possibility that time reversal symmetry is also broken, which further narrows down the order parameter that characterizes the hidden order. 

\bigskip

\end{abstract}

\section{Introduction}

The heavy fermion compound URu$_2$Si$_2$ possesses several different phases at low temperatures. At ambient pressure, two phase transitions at $T_{\rm HO}=17.5$\,K and $T_{\rm SC}=1.4$\,K have been found \cite{Pal85,Map86,Sch86}, the latter of which is the superconducting (SC) transition. The nature of the former transition has been a long-standing mystery, despite intense studies both experimentally and theoretically, and thus it is called hidden order (HO) transition \cite{Myd11}. Under high pressure, the HO phase transforms to the antiferromagnetic (AFM) phase with large staggered moment along the $c$ axis through the first-order phase transition, and the low-temperature SC phase disappears inside the AFM phase. 

In most cases (except for e.g. topological transitions), a phase transition accompanies symmetry breaking, from which the order parameter that characterizes the ordered phase arises. Without knowing which symmetries are broken in the ordered phase, many theoretical proposals of the HO parameter for URu$_2$Si$_2$ have been made \cite{Myd11,Kis05,Var06,Hau09,Cri09,Har10,Dub11}. Therefore the determination of broken symmetries below the transition temperature is essential to clarify the nature of the order. 

The broken symmetries in URu$_2$Si$_2$ are clearly identified in the AFM phase under pressure; in addition to the time reversal symmetry, translational symmetry is broken with the wave vector $\bm{Q}_{\rm C}=(0\,0\,1)$, and thus the Brillouin zone (BZ) is folded from the body-centered tetragonal space group $I4/mmm$ to the simple tetragonal $P4/mmm$. 

In the SC phase, the thermal conductivity \cite{Kas07,Kas09} and specific heat \cite{Yan08} measurements have revealed peculiar differences in the magnetic field dependence between the directions parallel and perpendicular to the $c$ axis, providing evidence for the presence of the point nodes along the $k_z$ direction in the superconducting energy gap. From this, an unconventional superconducting state with the chiral $d$-wave symmetry has been inferred \cite{Kas07,Kas09}, which has the superconducting gap with a form of 
\begin{equation}
\sin\frac{k_z}{2}c\left(\sin\frac{k_x+k_y}{2}a\pm\mathrm{i}\sin\frac{k_x-k_y}{2}a\right).
\label{chiral}
\end{equation}
This chiral $d$-wave symmetry is seemingly quite unusual, but it is consistent with the well-studied superconductivity mediated by the antiferromgantic spin fluctuations \cite{Kus11,Shi12}. This chiral superconducting state breaks time reversal symmetry, which has been a subject of recent studies including the lower critical field $H_{c1}$ \cite{Oka10}, high-field magnetic torque \cite{Li13} and Kerr effect measurements \cite{Kap14}. Among these, the temperature dependence of $H_{c1}$ shows an anomalous kink at $\sim 1.2$\,K, which is slightly lower than $T_{\rm SC}=1.4$\,K, only for the field parallel to the $c$ axis. Although the nature of this kink anomaly is not fully understood, a possibility of the separation of the two transition temperatures $T_{\rm SC1}=1.4$\,K and  $T_{\rm SC2}=1.2$\,K of the two states with the bases of $k_z(k_x+k_y)$ and $k_z(k_x-k_y)$ has been proposed, in which the time reversal broken (chiral) state with a complex order parameter $k_z(k_x+k_y)\pm {\rm i}k_z(k_x-k_y)$ appears only below the lower temperature $T_{\rm SC2}$. If this is correct, this immediately indicates that $k_x+k_y$ and $k_x-k_y$ are inequivalent in the normal state above $T_{\rm SC1}$, which means that the tetragonal rotational symmetry is broken in the HO phase. 

In this article, we review the current experimental situation of our understanding of broken symmetries in the HO phase of URu$_2$Si$_2$. We focus on three symmetries, (1) translational symmetry, (2) fourfold (tetragonal) rotational symmetry, and (3) time reversal symmetry.  The broken symmetries give strong constraints on the order parameter. In particular, the rotational symmetry breaking restricts the HO symmetry to the degenerate $E$-type irreducible representation \cite{Tha11}. Additional time reversal symmetry breaking pins down the symmetry as $E^-$ type. Several current theories consistent with these results are also reviewed. 

\section{Translational symmetry breaking}

\begin{figure}[t]
\begin{center}
\includegraphics[width=1.0\linewidth]{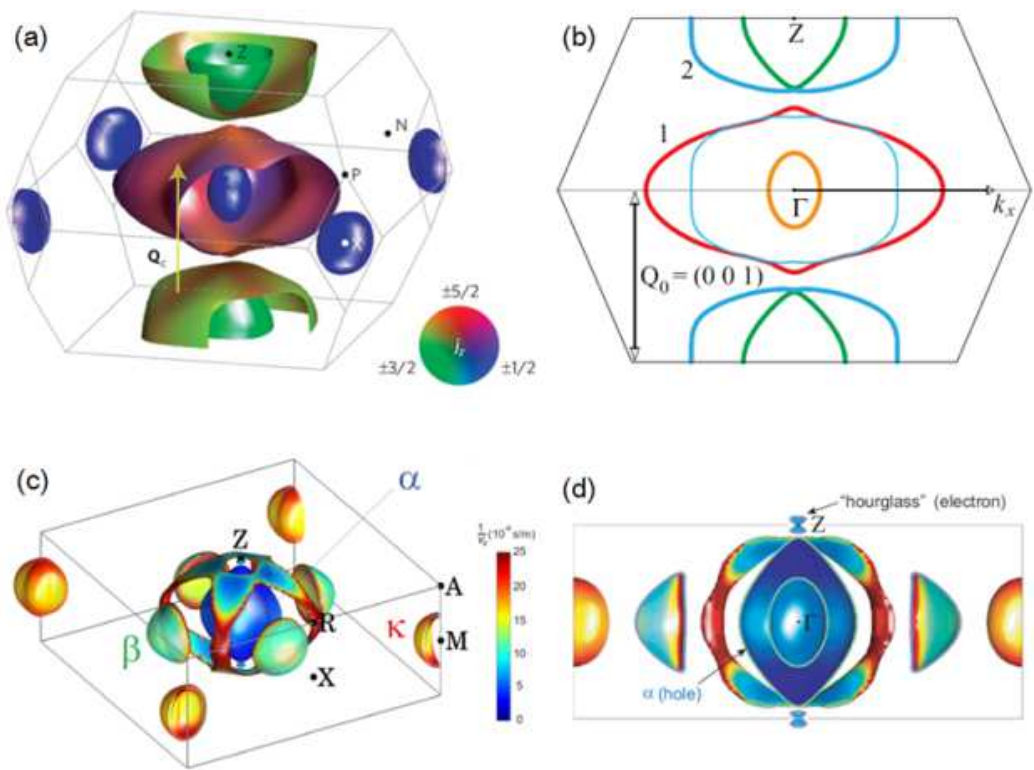}
\end{center}
\caption{Fermi surface structures in URu$_2$Si$_2$ calculated by the density-functional theory. (a) Paramagnetic FS colored by weight of the $j_z$ component \cite{Ike12}. (b) Cross sectional view of paramagnetic FS in a $k_xk_z$ plane including $\Gamma$ and $Z$ points \cite{Opp11}. Thin blue line is a shifted FS of band 2 (thick blue line) by the wave vector $\bm{Q}_{\rm C}=(0\,0\,1)$. (c) Three-dimensional view of FS assuming antiferromagnetism with a gap of 40\,meV, which corresponds to $\sim 4$\,meV when considering the renormalization factor of $\sim10$ \cite{Ton14}. The cage structure between the $\alpha$ band in the middle and four hemispherical $\beta$ pockets, which is one of the remnants of the nesting between bands 1 and 2 in (b), is sensitive to the gap value and can be easily diminished with a larger gap. Here the color corresponds to inverse of Fermi velocity $1/v_F$. (d) Cross sectional view of antiferromagnetic FS \cite{Ton14}. Thick lines correspond to the FS in the $k_xk_z$ plane including $\Gamma$ and $Z$ points. 
}
\label{FS}
\end{figure}

\begin{figure}[t]
\begin{center}
\includegraphics[width=\linewidth]{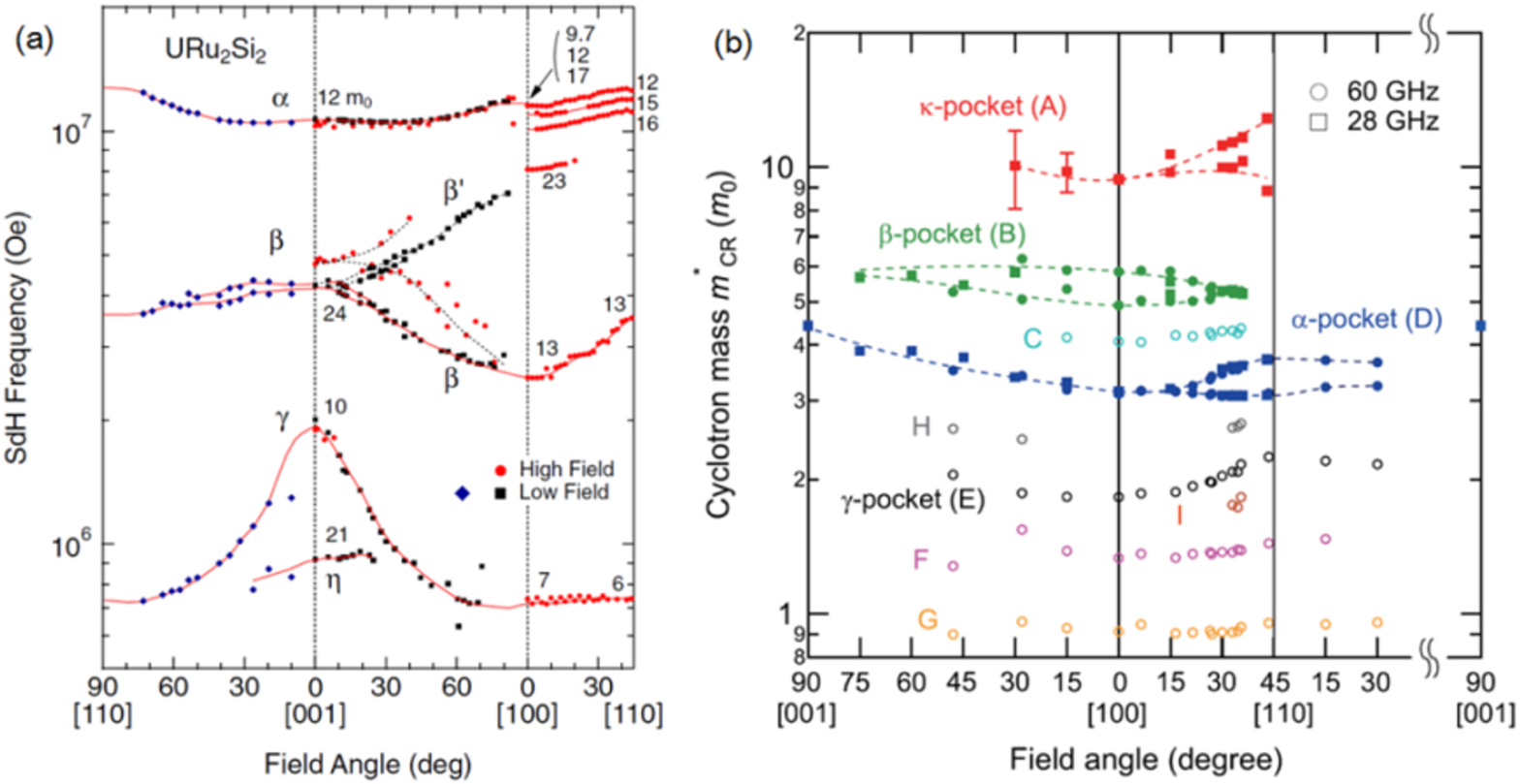}
\end{center}
\caption{Comparison between the field angle dependence of SdH frequency \cite{Aok12} (a) and cyclotron mass determined by the cyclotron resonance \cite{Ton14} (b). For the band assignments of $\alpha$, $\beta$ and $\kappa$, see Fig.\,\ref{FS}(c). The three branches with solid symbols guided by the dashed line in (b) are cyclotron resonance lines with strong intensities, which should correspond to the main bands ($\alpha$, $\beta$ and $\kappa$) with large FS volumes. Note that the cyclotron mass $m^*_{\rm CR}$ is different from the thermodynamic mass \cite{Ton14}.
}
\label{CR}
\end{figure}

First, we discuss the translational symmetry breaking in the HO phase. The band structure calculations in the paramagnetic phase by the density-functional theory (DFT) within the itinerant electrons picture \cite{Opp11,Ike12} (see Fig.\,\ref{FS}) show a good nesting between the largest electron Fermi surface (FS) sheet (band 1) and the largest hole FS sheet (band 2), which is characterized by the commensurate AFM wave vector $\bm{Q}_{\rm C}$ (see Fig.\,\ref{FS}(b) \cite{Opp11}).  The nested parts of the FS are dominated by the $j_z=\pm5/2$ component  (see Fig.\,\ref{FS}(a) \cite{Ike12}). Thus the FS structure suggests that URu$_2$Si$_2$ inherently exhibits the instability to antiferroic orderings characterized by $\bm{Q}_{\rm C}$. Indeed, in the pressure-induced AFM phase, the large moment appears along the $c$ axis, which doubles the primitive cell volume accompanying the band folding along $k_z$. Correspondingly, the $\Gamma$ and $Z$ points of the original body-centered tetragonal BZ become equivalent and transform to the $\Gamma$ point in the new simple tetragonal BZ (see Figs.\,\ref{FS}(c) and (d)). The nested parts of FS are gapped in the AFM phase, leading to the strong reduction in carrier numbers. 

In the HO phase, a large decrease in carrier concentrations has been indicated from the transport measurements \cite{Beh05,Kas07}, implying a gap formation in large parts of FS. The measurements of Shubnikov de Haas (SdH) effect under pressure \cite{Has10} have revealed no significant change in the quantum oscillation frequencies between the HO and AFM phases, which suggests that the FS structure in the HO phase is quite similar to that in the AFM phase. Moreover, the field angle dependence of one of the quantum oscillation frequency branches ($\beta$ in Fig.\,\ref{CR}(a)) shows a peculiar splitting ($\beta$ and $\beta$') for the field rotation from $[001]$ to $[100]$ directions \cite{Ohk99,Aok12}, which is consistent with the shape of four hemispherical electron $\beta$ pockets present in the FS structure in the AFM phase (see Figs.\,\ref{FS}(c) and (d)). Cyclotron resonance measurements \cite{Ton12,Ton14} have also revealed the characteristic angle dependence of the effective cyclotron mass branch corresponding to the $\beta$ pockets, which shows splitting behaviors for the field rotations from $[110]$ and $[001]$ towards the $[100]$ direction  (see $\beta$ in Fig.\,\ref{CR}(b)). Thus these bulk measurements under magnetic fields probing the electronic structure consistently indicate the presence of hemispherical pockets, which is in good agreement with the band structure calculations assuming the $\bm{Q}_{\rm C}$ band folding.

Recent angle resolved photoemission spectroscopy (ARPES) measurements also provide further evidence for the translational symmetry breaking in the HO phase \cite{Yos10,Boa13,Yos13,Men13}. The electronic structures near the $\Gamma$ and $Z$ points of the paramagnetic phase become similar below the HO transition temperature, indicating the band folding along $k_z$ \cite{Yos10,Boa13,Yos13}. The missing ARPES intensity along the $\Gamma$-$M$ line in the simple-tetragonal BZ (see Figs.\,\ref{FS}(c) and (d)) provides evidence for the gapping of the nested parts of FS in the paramagnetic phase \cite{Men13,San13}. Furthermore, the propeller-like crossing of the two orthogonal ellipsoids has been observed in the $M$ point, which also supports the $\bm{Q}_{\rm C}$ band folding \cite{Men13,San13}.

Thus, experimentally the translational symmetry breaking appears to be firmly established, and the HO parameter should be accompanied by the $\bm{Q}_{\rm C}$ band folding with gap formation in the nested parts of FS.

\section{Rotational symmetry breaking}

Next we discuss the rotational symmetry breaking in the HO phase of URu$_2$Si$_2$. Currently, four different experiments, magnetic torque \cite{Oka11}, cyclotron resonance \cite{Ton12,Ton14}, nuclear magnetic resonance (NMR) \cite{Kam13} and high-resolution X-ray scattering \cite{Ton13}, consistently have provided evidence for the in-plane twofold anisotropy, which breaks the tetragonal fourfold rotational symmetry.

\subsection{Magnetic torque}

\begin{figure}[t]
\begin{center}
\includegraphics[width=0.8\linewidth]{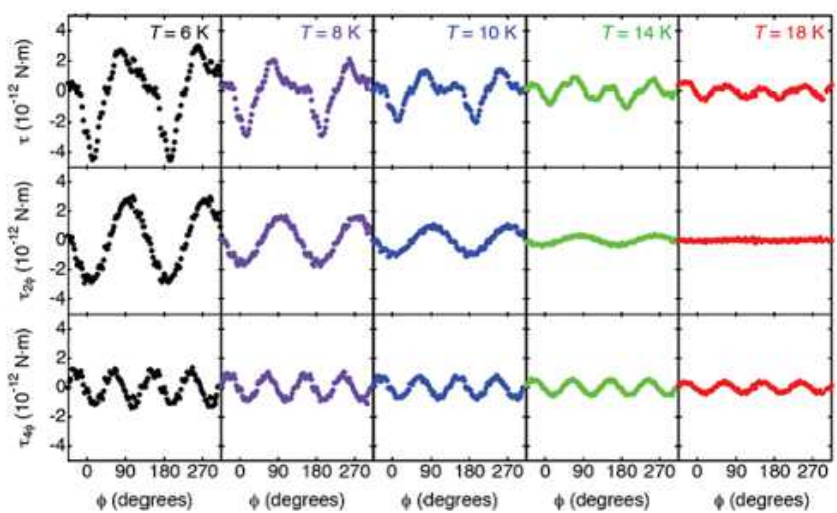}
\end{center}
\caption{In-plane magnetic torque above (red circles) and below $T_{\rm HO}=17.5$\,K (black, purple, blue and green circles) \cite{Oka11}. Upper panels show raw magnetic torque curves as a function of the azimuthal angle $\phi$ at several temperatures. All data are measured at $\left|\mu_0\bm{H}\right| = 4$\,T. Middle and lower panels show twofold $\tau_{2\phi}$ and fourfold $\tau_{4\phi}$ components of the torque curves which are obtained from the Fourier analysis.
}
\label{torque}
\end{figure}

In 2011, Okazaki {\it et al.} have reported the magnetic torque data under in-plane field rotation in small pure single crystals of URu$_2$Si$_2$ \cite{Oka11}. They found that the azimuthal angle $\phi$ dependence of the magnetic torque $\tau$ changes its shape at low temperatures below the HO transition at $T_{\rm HO}=17.5$\,K. $\tau(\phi)$ can be decomposed as $\tau=\tau_{2\phi}+\tau_{4\phi}+\tau_{6\phi}+\cdots$, where $\tau_{2n\phi}=A_{2n}\sin 2n(\phi-\phi_0)$ is a term with $2n$-fold symmetry with $n=1,2,\cdots$. The data for raw $\tau(\phi)$, twofold $\tau_{2\phi}$, and fourfold $\tau_{4\phi}$ components are shown in Fig.\,\ref{torque}. Above the HO transition, the torque is dominated by the fourfold term, and the twofold component is absent, which is consistent with the tetragonal symmetry of the crystal structure. In contrast, the twofold term is clearly observed below the transition, which is not expected in the system with the tetragonal symmetry. As the torque is a quantity given by the angle derivative of the free energy, this observation of the twofold term provides thermodynamic evidence for the fourfold rotational symmetry breaking. 

The $\phi$ dependence of the twofold component $\tau_{2\phi}$ gives the phase $\phi_0=45^\circ$, which indicates that the off diagonal susceptibility component $\chi_{ab}$ becomes finite below $T_{\rm HO}$. In other words, the in-plane twofold anisotropy of magnetic susceptibility elongated along the $[110]$ direction sets in below the transition. In such a state, electron fluid has a unidirectional response in analogy to a nematic state \cite{Fra10}. This kind of electronic nematic state is expected to have degeneracy, which leads to the formation of domains elongated along the $[110]$ and $[\bar{1}10]$ directions. In agreement with this picture, the relative magnitude of the twofold term $A_{2}$ has been found to become smaller for samples with larger size, in which the out-of-phase signals of $\tau_{2\phi}$ from the different nematic domains tend to cancel each other. The size of the domains is suggested to be of the order of 10\,$\mu$m \cite{Oka11}. This micro-domain formation would make it difficult to observe the in-plane anisotropy in macroscopic measurements such as transport etc. 

\subsection{Nuclear magnetic resonance (NMR)}

\begin{figure}[t]
\begin{center}
\includegraphics[width=0.7\linewidth]{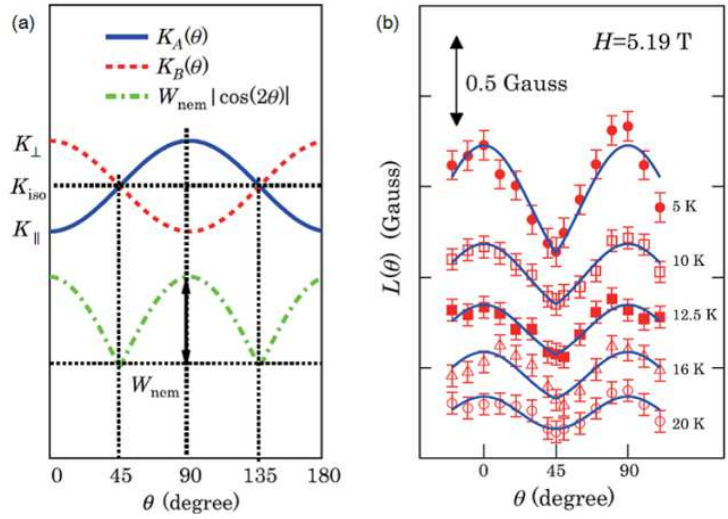}
\end{center}
\caption{NMR line width under in-plane field rotation \cite{Kam13}. Here the in-plane field angle $\theta$ is defined from $[110]$ direction, which has a relation to the azimuthal angle $\phi$ defined in the torque experiments \cite{Oka11} as $\theta=\phi-45^\circ$. (a) Schematic field angle dependence of NMR Knight shift expected for a state with broken fourfold symmetry. Blue solid and red dashed curves are the angluar dependences of Knight shift for two domains A and B, respectively. The averaged Knight shift is expected to be angle independent, but the corresponding line width (green dashed-dotted line) has a characteristic angle dependence with cusps. (b) $^{29}$Si NMR line width $L$ under in-plane field rotation at several temperatures. The solid lines are the fits to the angle dependence of line width taking into account the domain formation \cite{Kam13}. Each curve is shifted vertically for clarity.
}
\label{NMRangle}
\end{figure}

Another piece of evidence for the magnetic twofold anisotropy has been obtained by the NMR measurements, which can probe microscopic information on the magnetic susceptibility in terms of the Knight shift $K$. Kambe {\it et al.} measured the angular dependence of the $^{29}$Si NMR line width $L$ under in-plane field rotation \cite{Kam13}. They used a $^{29}$Si enriched sample to obtain high-resolution data. At high temperatures above the HO transition, they found sinusoidal angle dependence of $L$ consistent with the tetragonal symmetry, but in the HO state below $T_{\rm HO}$ they observed a characteristic cusp along the $[100]$ direction (see Fig.\,\ref{NMRangle}(b)). Such a cusp behavior can be explained by the line broadening associated with a superposition of the two sinusoidal curves of $K(\phi)$ having twofold symmetry (see Fig.\,\ref{NMRangle}(a)), which is fully consistent with the two nematic domains along the $[110]$ and $[\bar{1}10]$ directions inferred from the magnetic torque experiments \cite{Oka11}. Thus this gives microscopic evidence for the rotational symmetry breaking.

They have also made a quantitative estimate of the in-plane anisotropy of the Knight shift, which yields $\{\chi(H\parallel [110])-\chi(H\parallel [\bar{1}10])\}/\chi(H\parallel [100])= 2\chi_{ab}/\chi_{aa}=4\times 10^{-3}$, which is an order of magnitude smaller than the bulk estimate from the torque measurements. A possible reason of this discrepancy is that the NMR estimate assumes that the additional line width below the HO transition is purely paramagnetic ignoring possible internal field contributions associated with the ordering. In fact, as pointed out in a recent theory based on the antiferro-octupole order taking into account the broken fourfold rotational symmetry \cite{Han12}, for example, the NMR line width at Si site can exhibit a significant broadening due to the internal field associated with the multipole ordering at U site through the hyperfine interaction. If we consider the domain formation in such a state, we can show that such internal field contributions to the line width do not give a large difference between $H\parallel [100]$ and $H\parallel [110]$ configurations, which results in a nearly angle-independent broadening of the line width. In contrast, the bulk susceptibility can be anisotropic even in such a multipole state as in the case of an antiferromagntic state, in which the susceptibility along the staggered moment direction decreases towards zero in the zero temperature limit in contrast to the constant susceptibility for perpendicular directions. This difference may be an important clue to understand the nature of the HO, and deserves further studies.

\subsection{Cyclotron resonance}

An important information on an ordered phase can be obtained from the electronic structure, especially from the detailed structure of the FS. As described in Section 2, the HO phase has a similar FS structure with that in the pressure-induced AFM phase, having the $\bm{Q}_{\rm C}=(0\,0\,1)$ band folding. However, the HO and AFM phases are separated by the phase transition in the pressure phase diagram, and therefore identifying the characteristic feature of the electronic structure in the HO phase distinct from that in the AFM phase is even more important. 

The observation of cyclotron resonance \cite{Ton12,Ton14} owing to the high-quality single crystals with large mean free path \cite{Mat11} allowed the full determination of the angle dependence of the cyclotron masses for the main FS sheets (see Fig.\,\ref{CR}(b)), including the missing heavy FS in the quantum oscillation measurements which is now assigned as $\kappa$ band located around the zone corner (see Fig.\,\ref{FS}(c)). The three main bands, the $\alpha$ hole pocket around the $\Gamma$ point with the largest volume, the $\beta$ hemispherical electron pockets, and the heaviest $\kappa$ electron pockets, have characteristic angle dependences of effective mass. These dependences are mostly corresponding to their FS shapes obtained in the band-structure calculations in the AFM phase. However, a significant difference from the AFM calculations is found in the $\alpha$ branch near the $[110]$ direction. The cyclotron mass $m^*_{\rm CR}$ of $\alpha$ branch shows an anomalous splitting near the $[110]$ direction (see Fig.\,\ref{CR}(b)) in sharp contrast to the AFM calculations, in which nearly isotropic dependence is expected. Then it is important to understand this difference, which may involve a key information on the peculiar feature of the electronic structure in the HO phase.

The corresponding quantum oscillation frequency, which is a measure of the FS cross sectional area, exhibits a completely different three-fold splitting of the $\alpha$ branch with a constant separation in the entire field angle range within the $ab$ plane \cite{Aok12} (see Fig.\,\ref{CR}(a)). This three-fold splitting in the quantum oscillation frequency of the $\alpha$ branch, which disappears with a small field tilt from the $ab$ plane, has been suggested to be originate from the magnetic breakdown effect \cite{Ton14} between the $\alpha$ hole pocket and the nearby small electron pockets around the $Z$ points with an hourglass shape (see Fig.\,\ref{FS}(d)). It is therefore plausible that the shape of the $\alpha$ hole FS is essentially isotropic within the plane. The cyclotron mass split near the $[110]$ direction observed in the cyclotron resonance at much lower fields, which is a robust feature against the field tilt, should have a different origin from the magnetic breakdown. It can be rather naturally understood by the twofold symmetry of the in-plane mass anisotropy with taking into account the domain formation. As discussed by Tonegawa {\it et al.} \cite{Ton12,Ton14}, the broken fourfold symmetry can lead to the imbalance of the interband scattering between the corners of the $\alpha$ and $\beta$ FS pockets which are connected with the incommensurate wave vector $\bm{Q}_{\rm IC}\approx(0.4\,0\,0)$. This will make the in-plane mass anisotropic having hot spots along the $[110]$ direction, which can explain the observed splitting when considering the presence of two domains. In this case, the mass enhancement at the hot spots is due to electron correlation effects, which may not change the FS shape significantly, consistent with the quantum oscillation results. We also note that the information on the effective mass can also be obtained from quantum oscillations, but this requires the fitting of the temperature dependence of the oscillation amplitude to the Lifshitz-Kosevich formula, which may be difficult to resolve two slightly different masses. 

Thus we have now evidence for the broken fourfold rotational symmetry in the HO phase from the electronic structure. The cyclotron resonance measurements reveal the in-plane mass anisotropy, which breaks the tetragonal symmetry.  

\subsection{High-resolution X-ray scattering}

The above three experiments, magnetic torque, NMR, and cyclotron resonance, have been all done in the presence of in-plane magnetic field. As the anomalies are found in the angle dependence under the in-plane field rotation, the broken fourfold symmetry in these measurements are not resulting from the presence of symmetry breaking field. However, the magnetic field can induce additional order which may complicate the discussion regarding the nature of the HO in the ground state at zero magnetic field. It is therefore important to obtain experimental evidence in the absence of magnetic field. 

Quite recently, a high-resolution crystal structure analysis has been performed at zero magnetic field by using synchrotron X-ray scattering at SPring-8. By focusing on the high-angle $(880)$ Bragg peak, which is sensitive to the orthorhombic distortion associated with the twofold symmetry inferred from the above experiments, Tonegawa {\it et al.} have succeeded in observing a clear peak split below the HO transition in a very clean crystal of URu$_2$Si$_2$ with a high residual resistivity ratio $RRR\sim 670$ \cite{Ton13}. They tune the X-ray energy so that the attenuation length is as long as $\sim32\,\mu$m, which is expected to be larger than the domain size. The observed peak splitting clearly indicates the domain formation due to the orthorhombic distortion. The crystal structure in the HO phase is found to belong to the $Fmmm$-type orthorhombic symmetry which breaks the fourfold rotational symmetry and the $ab$ primitive vectors are rotated $45^\circ$ from the tetragonal phase. This symmetry is fully consistent with the twofold in-plane anisotropy found in the torque, NMR and cyclotron resonance experiments under in-plane field rotation. Thus this is the first zero-field evidence of the broken tetragonal symmetry from the scattering experiments, which strongly suggests that the electronic in-plane anisotropy found in these measurements is not field induced, but is an intrinsic property of the HO phase. 

The observed orthorhombicity $\delta=(a-b)/(a+b)$ is of the order of $10^{-5}$, which is much smaller than that in the orthorhombic phase of BaFe$_2$As$_2$-based superconductors \cite{Kas12} with an isomorphic crystal structure. This smallness implies that the HO transition is driven by the electronic ordering, which distorts the lattice owing to the small electron-lattice coupling. Indeed, this is consistent with the fact that the HO is likely an antiferroic order with $\bm{Q}_{\rm C}$ band folding, which should couple only weakly to the ferroic orthorhombicity. 

We note that previous experiments focusing on $(h00)$ Bragg peaks using much dirtier samples with $RRR\sim 10$ have not shown any orthorhombicity in the HO phase \cite{Nik10,Bou11}. In agreement with this, the $(880)$ Bragg peak splitting has been observed only in the very clean crystal, which suggests that the impurity scattering masks the coupling between the electronic nematic order and the lattice distortion \cite{Ton13}. Such an unusual impurity effect of nematic order deserves further studies. 

\section{Time reversal symmetry breaking}

The next important issue is whether time reversal symmetry is broken or not in the HO phase. Direct evidence for or against time reversal symmetry breaking has not been obtained to date. In the early neutron scattering experiments \cite{Bro87}, a small antiferromagnetic moment along the $c$ axis has been reported, but later this has been found to be due to some inhomogeneity in the sample and no intrinsic moment along the $c$ axis has been identified in the HO phase \cite{Mat01,Tak07}. It should be noted that the experimental evidence for the rotational symmetry breaking in the $ab$ plane discussed in Section 3 leads to a search for the in-plane staggered moment. However, such a moment has never been reported in the neutron scattering measurements so far within the experimental resolution of $\sim 10^{-2}\mu_B$. 

\begin{figure}[t]
\begin{center}
\includegraphics[width=\linewidth]{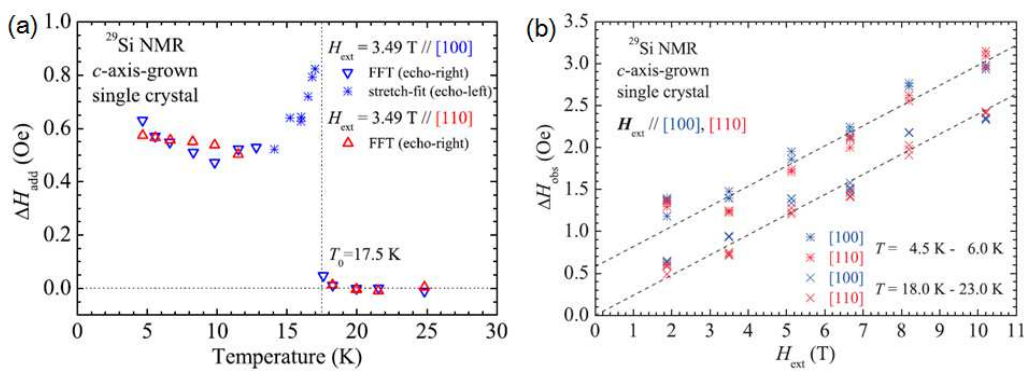}
\end{center}
\caption{$^{29}$Si NMR line width broadening with in-plane magnetic fields \cite{Tak11}. (a) Temperature dependence of the additional broadening $\Delta H_{\rm add}$ for $H \parallel [100]$ (blue) and for $H \parallel [110]$ (red). A constant width observed above the HO transition has been subtracted. The vertical dotted line marks the HO transition temperature $T_{\rm HO}$. (b) Field dependence of  $\Delta H_{\rm add}$ below and above $T_{\rm HO}=17.5$\,K. 
The dashed lines are the fits to the linear $H$ dependence.}
\label{NMRfield}
\end{figure}

A piece of evidence for the broken time reversal symmetry has been inferred from the recent NMR analysis. Takagi {\it et al.} measured $^{29}$Si NMR with in-plane magnetic fields and found that additional broadening of the spectrum $\Delta H_{\rm add}$ sets in just below $T_{\rm HO}$ \cite{Tak11} (see Fig.\,\ref{NMRfield}(a)). The additional broadening $\Delta H_{\rm add}$ has unusual temperature dependence with a jump at $T_{\rm HO}$ and decreases just below $T_{\rm HO}$. Such a broadening is consistent with the hyperfine internal field at Si site generated from the U-site multipole ordering with time reversal symmetry breaking \cite{Han12}. We note that similar temperature dependence is found in the orthorhombicity $\delta(T)$ in the synchrotron X-ray measurements introduced in Section 3.4 \cite{Ton13}, suggesting a close correspondence between the hyperfine field and the orthorhombic distortion. 

The field dependence of $\Delta H_{\rm add}$ shows a roughly linear-in-$H$ dependence (see Fig.\,\ref{NMRfield}(b)), and most importantly, it intersects the $H=0$ axis at a finite value of $\sim 0.6$\,Oe. This finite $\Delta H_{\rm add}(H\to 0)$ has been interpreted in terms of the broken time reversal symmetry in the HO phase. The obtained value of the additional broadening would correspond to the actual in-plane antiferromagnetic moment 
much smaller than the neutron resolution. 

It is very important to establish the time reversal symmetry is really broken in the HO state by accumulating further experimental evidence by using several other measurement techniques, such as muon spin rotation \cite{Kaw13} and Kerr rotation \cite{Kap14}. 

\section{Constraints on the symmetry of the hidden order parameter}

Broken symmetries give stringent constraints on the symmetry of the order parameter. The broken translational symmetry indicates that the order is an antiferroic type with the wave vector given by $\bm{Q}_{\rm C}=(0\,0\,1)=(1\,0\,0)$. This rules out the possible HO parameters associated with the incommensurate wave vector $\bm{Q}_{\rm IC}$, although it has been suggested by the inelastic neutron measurements that the spin excitation gap at $\bm{Q}_{\rm IC}$ is related to the large specific heat jump at $T_{\rm HO}$ \cite{Wie07}.

The fourfold rotational symmetry breaking acts as a further strong constraint on the order parameter. Thalmeier and Takimoto \cite{Tha11} have shown from the Landau free energy analysis of the multipole ordering that among the allowed irreducible representations (four non-degenerate $A_1$, $A_2$, $B_1$, $B_4$ and one degenerate $E$ symmetries), the observed twofold signals in the magnetic torque \cite{Oka11} are consistent only with the $E$ type. When considering the parity ($+$ or $-$) with respect to time reversal, this $E$ type can be divided into $E^+$ and $E^-$ symmetries \cite{Ike12,Tha13}. Within the multipole orders belonging to the $E$ type, even rank quadrupole (rank-2) \cite{Tha11} and hexadecapole (rank-4) $E^+$ orders with preserved time reversal symmetry, and odd rank octupole (rank-3) \cite{Han12} and dotoriakontapole (rank-5) $E^-$ \cite{Ike12,Rau12} orders with broken time reversal symmetry are the candidates of the HO parameter consistent with the rotational symmetry breaking. Recent first-principle multipole calculations taking into account for the vertex corrections by Ikeda {\it et al.} find that among the all allowed symmetries, the rank-5 $E^-$ order exhibits the largest susceptibility at the $\bm{Q}_{\rm C}$ wave vector, indicating that this dotoriacontapole is the first candidate of the HO parameter. It has been also shown theoretically \cite{Rau12} that this state indeed has a finite $\chi_{ab}$ component, consistent with the nematic response found in the torque experiments. 

More exotic states than such multipole approaches, which break the rotational symmetry, have also proposed theoretically. An example for time reversal preserving state is the spin nematic order \cite{Fuj11,Ris12}. A difference from the multipole $E^+$ orders is that this spin nematic order has a momentum dependent order parameter \cite{Fuj11}, which may have nodes for certain directions in analogy to e.g. $d$-wave superconductivity. Another example for broken time reversal symmetry is the so-called hastatic order proposed recently by Chandra, Flint and Coleman \cite{Cha13}. This state is also anisotropic and the V shaped density of states has been suggested. 

\section{Summary}

We have reviewed the current status of our understanding of broken symmetries in the mysterious HO phase of the heavy fermion compound URu$_{2}$Si$_{2}$. Experimental evidence for translational symmetry breaking is firmly established. Thus the HO parameter should be antiferroic and have the commensurate wave vector $\bm{Q}_{\rm C}$. There is growing evidence for the rotational symmetry breaking, and magnetic susceptibility anisotropy is elongated along the $[110]$ direction, which has been consistently found in the magnetic torque and NMR measurements. The cyclotron resonance experiments have provided the electronic structure evidence that the FS breaks the fourfold rotational symmetry. Recent high-resolution synchrotron X-ray scattering experiments in very clean sample have revealed a small lattice change from tetragonal to orthorhombic structure, giving direct evidence from crystal structure for fourfold rotational symmetry breaking at zero magnetic field. These results restrict the symmetry of the HO parameter to belonging to the $E$-type irreducible representation. 

To pin down the genuine order parameter, the next important step is to establish the parity with respect to time reversal, which has been suggested to be odd from an NMR analysis. Further experimental studies are necessary to clarify this point. Another point which is helpful is whether the order parameter has momentum dependence or not. This requires momentum resolved spectroscopic techniques which are sensitive to low energy excitations, such as high-resolution ARPES and quasiparticle interference by using scanning tunneling microscopes. 

\section*{Acknowledgments}

This review is based on the work in collaboration with T. Fukuda, Y. Haga, K. Hashimoto, K. Ikada, S. Kasahara, Y.-H. Lin, T.\,D. Matsuda, Y. Mizukami, R. Okazaki, Y. Onuki, H.\,J. Shi, H. Shishido, K. Sugimoto, S. Tonegawa, Y. Tsuruhara, D. Watanabe, E. Yamamoto, and N. Yasuda. We thank D. Aoki, A.\,V. Balatsky, K. Behnia, F. Bourdarot, P. Chandra, P. Coleman, R. Flint, S. Fujimoto, K. Ishida, H. Harima, S. Kambe, A. Kaptulnik, Y. Kasahara, I. Kawasaki, G. Knebel, G. Kotliar, Y. Kuramoto, H. Kusunose, J.\,A. Mydosh, P.\,M. Oppeneer, P.\,S. Riseborough, A. F. Santander-Syro, M. Sigrist, T. Takimoto, P. Thalmeier, C.\,M. Varma, H. Yamagami, Y. Yanase, and Y. Zaanen for helpful discussions. This work was supported by Grant-in-Aid for the Global COE program ``The Next Generation of Physics, Spun from  Universality and Emergence", Grants-in-Aid for Scientific Research on Innovative Areas ``Heavy Electrons'' (No.\,20102002, 20102006, 23102713)  and ``Topological Quantum Phenomena" (No.\,25103713) from the Ministry of Education, Culture, Sports, Science and Technology (MEXT) of Japan, and KAKENHI from the Japan Society for the Promotion of Science (JSPS).  

\end{document}